\documentclass[prl,twocolumn,amsmath,amssymb,superscriptaddress,floatfix]{revtex4}
\usepackage{epsf}
\usepackage{bm}

\begin{document}


\title{Pinning of dynamic spin density wave
fluctuations\\ in the cuprate superconductors}

\author{Anatoli Polkovnikov}
\affiliation{Department
of Physics, Yale University, P.O. Box 208120, New Haven CT
06520-8120}

\author{Matthias Vojta}
\affiliation{Theoretische Physik III, Elektronische Korrelationen
und Magnetismus, Universit\"at Augsburg, 86135 Augsburg, Germany}

\author{Subir Sachdev}
\affiliation{Department
of Physics, Yale University, P.O. Box 208120, New Haven CT
06520-8120}

\date{February 25, 2002}

\begin{abstract}
We present a theory of the pinning of dynamic spin density wave
(SDW) fluctuations in a $d$-wave superconductor by local
imperfections which preserve spin-rotation invariance, such as
impurities or vortex cores. The pinning leads to static spatial
modulations in spin-singlet observables, while the SDW
correlations remain dynamic: these are the `Friedel oscillations'
of a spin-gap antiferromagnet. We connect the spectrum of these
modulations as observed by scanning tunnelling microscopy to the
dynamic spin structure factor measured by inelastic neutron
scattering.
\end{abstract}

\maketitle

Many studies of the cuprates assume that high temperature
superconductivity is best understood by a theory of doping mobile
charge carriers into a Mott insulator which is paramagnetic
\cite{pwa} {\em i.e.} a Mott insulator with no static spin moment
on any site and dynamic antiferromagnetic spin density wave (SDW)
correlations. Studies of paramagnetic Mott insulators on the
square lattice of Cu ions \cite{vs} showed that the most promising
candidates have broken translational symmetry (and confinement of
$S=1/2$ spinon excitations) associated with the appearance of
spontaneous bond charge order. Here, and henceforth, we define the
term ``charge order'' very generally: there is a periodic spatial
modulation in observables which are invariant under spin rotations
and time reversal, such as local electron/hole density of states
(LDOS), the spin exchange energy or the electron kinetic energy
contained in a link of the square lattice; the modulation in the
total charge density may well be unobservably small because of
screening by the long-range Coulomb interactions.

Upon doping the charge-ordered paramagnetic Mott insulator
\cite{vs}, superconductivity co-exists with charge order for a
finite range of carrier concentrations ($\delta$), and a $d$-wave
superconductor with the full square lattice symmetry appears above
a critical $\delta$. Almost all theoretical and experimental
studies have concluded that the low temperature properties of this
$d$-wave superconductor at optimal $\delta$ are qualitatively
identical to those of a BCS superconductor obtained by the Cooper
instability of a metallic Fermi liquid. This conclusion raises the
question of whether the connection to the Mott insulator is even
necessary in a description of the low temperature properties of
the superconductor at or above optimal $\delta$.

A number of works \cite{vortex} argued that memory of the Mott
insulator should survive in and around vortices in superconducting
order: the suppression of superconductivity in the vortex core
implies that the Cooper pairs are not condensed, but the electrons
should still strive to retain the exchange correlation energy of
the Mott insulator. This reasoning, and the studies of the doped
paramagnetic Mott insulator \cite{vs}, led to the suggestion
\cite{ps} that the static charge order, along with dynamic SDW
correlations, should appear near vortex cores.

Hoffman {\em et al.} \cite{seamus} introduced a novel scanning
tunnelling microscopy (STM) technology of atomically registered
spectroscopic mapping. Applied to slightly over-doped
Bi$_2$Sr$_2$CaCu$_2$O$_{8+x}$, they detected LDOS modulations
around the vortex cores at wavevectors ${\bf K}_{cx} = (\pi/2,0)$
and ${\bf K}_{c y} = (0,\pi/2)$ (the square lattice spacing has
unit length and this modulation has a period of four lattice
spacings), co-existing with well established superconductivity
\cite{sdw}.
Bulk charge order with this, and related, periods has been
discussed in insulating/superconducting paired hole states from a
number of different theoretical perspectives
\cite{vs,white,jan,steve,troyer,ps,castro,assa}.
 Neutron scattering studies of the optimally doped cuprates
have not (so far) seen dynamic or static charge order, but do see
collinear SDW fluctuations at wavevectors ${\bf K}_{sx} = (3\pi
/4, \pi)$ and ${\bf K}_{sy} = (\pi, 3\pi /4)$ and at energies
above a spin gap $\Delta$. On symmetry grounds \cite{zachar},
collinear SDW correlations are accompanied by charge order
correlations at wavevectors $2 {\bf K}_{sx,y}$, which equal the
values of $\pm{\bf K}_{cx,y}$, modulo reciprocal lattice vectors.
This connection was used \cite{pphmf} to describe the spatial
envelope of the charge order in the doped Mott paramagnet by the
pinning of a `sliding' degree of freedom of the SDW (the
spin-rotation degree of freedom is not pinned and remains dynamic)
by imperfections, such as vortex cores.

This paper presents a description of the energy dependence of the
LDOS measured in STM in a simple Gaussian model of the pinning of
dynamic SDW fluctuations. We connect the spectroscopic information
contained in the STM observation at wavevectors ${\bf K}_{cx,y}$
to the inelastic neutron scattering spectra at wavevectors ${\bf
K}_{sx,y}$. Our theory applies to pinning by impurities and most
other imperfections that preserve spin rotation invariance. It
also applies to pinning by vortex cores, but neglects the Doppler
shift in the quasiparticle energies induced by the superflow
\cite{volovik}: this shift has no contribution at the wavevectors
${\bf K}_{cx,y}$, and so its neglect is justifiable. Our results
should be distinguished from other recent proposals \cite{ting}
which assume {\it static} SDW order, for which there is no direct
evidence in the system of Ref.~\onlinecite{seamus}.

We begin by identifying the complex order parameters,
$\Phi_{x\alpha}$, $\Phi_{y\alpha}$ ($\alpha=x,y,z$ extends over
the directions in spin space) for SDW fluctuations; the spin
operator on the lattice site ${\bf r}$ is
\begin{displaymath}
S_{\alpha} ({\bf r}, \tau) = \mbox{Re} \left[e^{i {\bf K}_{sx}
\cdot {\bf r}} \Phi_{x \alpha} ({\bf r}, \tau)+  e^{i {\bf K}_{sy}
\cdot {\bf r}} \Phi_{y \alpha}({\bf r}, \tau)\right],
\end{displaymath}
where $\tau$ is imaginary time. We use a Gaussian action for
$\Phi_{x,y\alpha}$ fluctuations:
\begin{equation}
\mathcal{S}_{\Phi} = T \sum_{{\bf q}, \omega_n, \alpha}
\chi^{-1}_x ( {\bf q}, \omega_n ) | \Phi_{x \alpha} ({\bf q},
\omega_n) |^2 + (x \rightarrow y);
\end{equation}
we have Fourier transformed from $({\bf r}, \tau)$ to wavevectors
${\bf q}$ and Matsubara frequencies $\omega_n$ at a temperature
$T$. The dynamic spin susceptibilities $\chi_{x,y}$ will be inputs
from which we will deduce the dynamic structure of the pinned
charge order. However, this should not be interpreted as a causal
connection from the SDW fluctuations to the charge order; rather,
our method relies on the assumed proximity of the superconductor
to a quantum transition to a superconducting state with long-range
SDW order \cite{vs,sdw}. The presence of such a quantum critical
point, and the values of ${\bf K}_{sx,y}$ near it, are determined
by a complicated interplay between spin and charge correlations in
the superconductor and the proximate Mott insulator. A
corresponding theory using the charge density wave order parameter
can also be developed near a charge-ordering transition in the
superconductor; this is not discussed here and it yields results
closely related to those in \cite{pphmf}. We will take a simple
functional form for $\chi_{x,y}$ (see below), with a spectral
weight which vanishes below a spin gap energy $\Delta$ (a more
accurate form can, in principle, be extracted from neutron
scattering experiments). The value of $\Delta$ can be tuned by an
applied magnetic field perpendicular to the square lattice layers
\cite{sdw}, and its primary effect is a reduction in $\Delta$.

Invariance under translation by $n_x$ lattice spacings in the $x$
direction (for arbitrary integer $n_x$) leads to the symmetry
$\Phi_{x,\alpha} \rightarrow e^{i n_x \pi /4} \Phi_{x \alpha}$
obeyed by $\mathcal{S}_{\Phi}$ (and similarly for
$\Phi_{y\alpha}$). As long as this symmetry is preserved, no
observable will exhibit static charge order at wavevectors ${\bf
K}_{cx,y}$. However, this sliding translation symmetry is broken
by any imperfection (vortex core or impurity) that may be centered
near ${\bf r} = {\bf r}_0$, provided the spatial extent of the
imperfection is not much larger than the charge order period (4
lattice spacings). We account for such an imperfection by adding a
pinning term to the action
\begin{equation}
\mathcal{S}_{\rm pin} = -\sum_{\alpha} \int d \tau \left[ \zeta_x
\Phi_{x \alpha}^2 ({\bf r}_0, \tau) + (x\rightarrow y) + c.c.
\right], \label{spin}
\end{equation}
where $\zeta_{x,y}$ are complex coupling constants representing
the pinning potential. Note that $\mathcal{S}_{\rm pin}$ contains
the simplest terms which break the $\Phi_{x,\alpha} \rightarrow
e^{i n_x \pi /4} \Phi_{x \alpha}$ symmetry, but preserve spin
rotation invariance. Additional terms like $\Phi_{x \alpha} ({\bf
r}_0, \tau) \Phi_{y \alpha} ({\bf r}_0, \tau)$ are also permitted
but we will not include them for simplicity; such terms lead to a
$90^{\circ}$ turn in an SDW fluctuation and can be expected to
have a small amplitude on physical grounds---their presence will
lead to additional charge order at $\pm {\bf K}_{sx} \pm {\bf
K}_{sy}$. We also note that the theory for the superconductor
proximate to a charge ordering transition has a pinning term
linear in the charge order parameter.

STM involves tunneling of single electrons, and so to compute the
STM spectra we couple the Gaussian SDW fluctuations described by
$\mathcal{S}_{\Phi} + \mathcal{S}_{\rm pin}$ to the electrons,
which we describe by a standard $d$-wave BCS model:
\begin{equation}
H_{\rm BCS} = \sum_{\bf k} \Psi^{\dagger}_{\bf k} \left[
(\varepsilon_{\bf k} - \mu) \tau^z + \Delta_{\bf k} \tau^x \right]
\Psi_{\bf k}.
\end{equation}
Here $\Psi_{\bf k} = (c_{{\bf k}\uparrow}, c_{-{\bf k}\downarrow}^{\dagger})$ is
a Nambu spinor at momentum ${\bf k}=(k_x, k_y)$,
$\tau^{x,y,z}$ are Pauli matrices in particle-hole space, and $\mu$ is
a chemical potential.
For the kinetic energy, $\varepsilon_{\bf k}$ we
have first ($t$) and second ($t^{\prime}$)
neighbor hopping, while we assume a
$d$-wave form for the BCS pairing function $\Delta_{\bf k} =
(\Delta_0/2)
(\cos k_x - \cos k_y )$.
The full Hamiltonian for the conduction electrons reads
\begin{eqnarray}
H &=& H_{BCS} + \gamma \sum_{{\bf r}} c^{\dagger}_{\mu} ({\bf r})
\sigma^{\alpha}_{\mu\nu} c_{\nu} ({\bf r}) \times \\
&\times& \left( \Phi_{x \alpha} ({\bf r}) e^{i {\bf K}_x \cdot
{\bf r}} + \Phi_{x \alpha}^{\ast} ({\bf r}) e^{-i {\bf K}_x \cdot
{\bf r}} + (x\rightarrow y)  \right), \nonumber
\end{eqnarray}
where $\sigma^{\alpha}$ are the Pauli spin matrices, and $\gamma$
describes the scattering of the Bogoliubov quasiparticles of the
superconductor off the SDW fluctuations.

\begin{figure}
\epsfxsize=3.5in \centerline{\epsffile{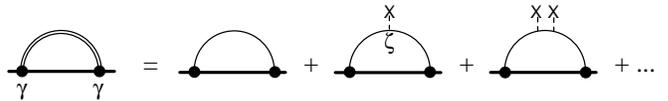}} \caption{
One-loop diagrams for the self energy of the host conduction
electrons. Thick lines are bare conduction electron propagators, a
single/double thin line denotes a bare/full SDW propagator, and
the cross represents the scattering of the SDW fluctuations by the
imperfection pinning term. } \label{figdgr}
\end{figure}

We computed the influence of the collective SDW modes on the
single particle properties in the one-loop approximation sketched
in Fig~\ref{figdgr} with processes of second order in the
electron-SDW coupling $\gamma$.
Because of the breaking of translational symmetry in
$\mathcal{S}_{\rm pin}$, the computations were numerically
involved: we used lattices of sizes up to 32$\times$32 with
periodic boundary conditions, and numerically inverted matrices in
real space to transform between the self energy and the Green's
function.

\begin{figure}
\epsfxsize=3.5in \centerline{\epsffile{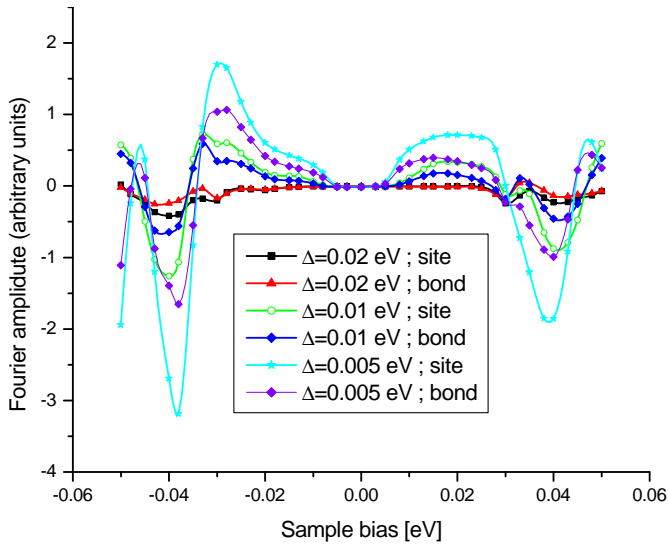}} \caption{The
$\zeta_{x,y} \gamma^2$ term of (\protect\ref{ft}), as a function
of the energy, $\omega$, of the electron. The BCS superconductor
is parameterized by hopping strengths
$t=-0.15$ eV, $t'=-t/4$, 
chemical potential $\mu = -0.135$ eV which gives a doping level of
about 15\%, and a size of the $d$-wave gap of $\Delta_0=40$~meV.
Results are shown for different values of $\Delta$, with $c=0.2$
eV and ${\bf r}_0$ on a square lattice site or bond (results for
${\bf r}_0$ on a plaquette are similar to those on a bond).}
\label{tolyaft}
\end{figure}

Our main results are apparent by considering the linear effect of
the lowest order diagram which breaks translational invariance:
this is the second diagram on the r.h.s. of Fig.~\ref{figdgr}, and
is of order $\zeta_{x,y}\gamma^2$. We computed its influence on
the LDOS in a theory with $J\chi_{x,y}^{-1} = \omega_n^2 + c^2
{\bf q}^2 + \Delta^2$ ($J$ is a characteristic spin fluctuation
energy scale, and $c$ is a spin-wave velocity) so that, after a
Fourier transform to real space, the $\Phi_{x,y}$ propagators
became Bessel functions of ${\bf r}$. Following the method of
Hoffman {\em et al.} \cite{seamus}, we computed the energy
($\omega$) dependence of
\begin{equation}
\sum_{{\bf r}} e^{-i {\bf K}_{cx,y} \cdot {\bf r}} \mbox{Im}
\left. c_{\mu} ({\bf r}, \omega_n) c_{\mu}^{\dagger} ({\bf r},
\omega_n) \right|_{i \omega_n \rightarrow \omega+i0^{+}},
\label{ft}
\end{equation}
the coefficients of the spatial Fourier transform of the LDOS at
wavevectors ${\bf K}_{cx,y}$. The $\zeta_{x,y} \gamma^2$ term in
(\ref{ft}) was evaluated by summations over ${\bf r}$ and
frequency space. Its values are shown in Fig~\ref{tolyaft} for a
fermion band corresponding to an optimally doped cuprate.
This method has the advantage that the values of $\gamma$, and
$\zeta_{x,y}$ only modify an overall pre-factor of the modulation,
and so do not effect the energy dependence of the results in
Fig~\ref{tolyaft}. The change in sign of the LDOS at $|\omega| <
\Delta_0$ arises from the dynamic nature of the SDW fluctuations,
and the resulting interference between the real and imaginary
parts of the bare propagator and the self energy.

We also performed computations in a {\em different} model in which
all the diagrams in Fig~\ref{figdgr} were summed. We introduced
two {\it real} three-component fields $\varphi_{x,y\alpha} ({\bf
r},\tau) = \mbox{Re} [e^{i {\bf K}_{sx,y} \cdot {\bf r}} \Phi_{x,y
\alpha} ({\bf r}, \tau)]$, with the $\varphi_{x,y\alpha}$ taking
momenta over the whole Brillouin zone of the square lattice --
this simplifies the exact treatment of the pinning term with a $T$
matrix. The bulk fluctuations were controlled by the
susceptibilities $\chi_{\varphi_{x,y}}({\bf q}, \omega_n) =
J/(\omega_n^2 + \omega_{x,y}({\bf q})^2)$, where the dispersion
$\omega_{x,y}({\bf q})^2 = \Delta^2 + c^2 \min(|{\bf q}+{\bf
K}_{sx,y}|,|{\bf q}-{\bf K}_{sx,y}|)^2$ \cite{dispnote}. An
important difference in this model is that the pinning term is now
$-\zeta \int d \tau [ \varphi_{x \alpha}^2 ({\bf r}_0, \tau) +
\varphi_{y \alpha}^2 ({\bf r}_0 , \tau)]$, with $\zeta$ real; in
terms of the previous formulation, this includes both
$|\Phi_{x,y\alpha}|^2$ and $\Phi_{x,y\alpha}^2$ terms with equal
modulus coefficients. This equality is a weakness of this second
model because \cite{sdw} the physical origin of the two terms is
very different, and $|\Phi_{x,y\alpha} ({\bf r}_0)|^2$ is better
accounted for by a shift in the local spin exciton frequency.

\begin{figure}
\epsfxsize=3.5in \centerline{\epsffile{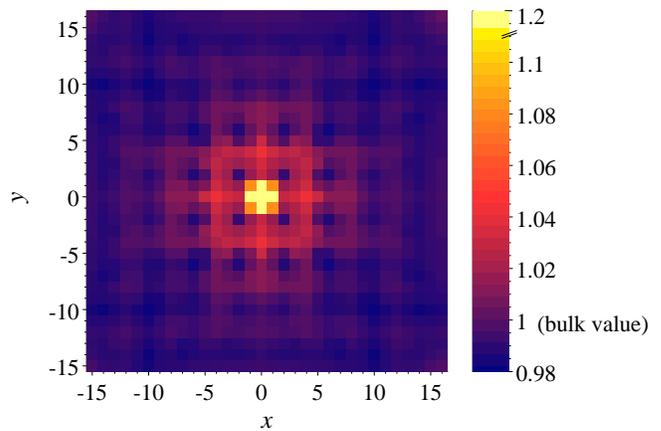}} \caption{
Grayscale plot of the quasiparticle LDOS near the imperfection at
lattice site ${\bf r}_0 = (0,0)$, measured at $\omega=-0.75
\Delta_0$. The parameters of $H_{BCS}$ are as in
Fig~\protect\ref{tolyaft}. The SDW fluctuations are characterized
by a gap of $\Delta=5$~meV, $J=0.15eV$ and a velocity of
$c=0.2$~eV; the SDW-fermion couplings is $\gamma = 0.1$ eV and the
pinning strength is $\zeta = 0.4$ eV. A similar modulation occurs
for all energies $-2 \Delta_0 < \omega < -\Delta_0/8$, and
significantly weaker at positive bias $\Delta_0/8 < \omega < 2
\Delta_0$. } \label{figldos1}
\end{figure}

\begin{figure}
\epsfxsize=3.5in \centerline{\epsffile{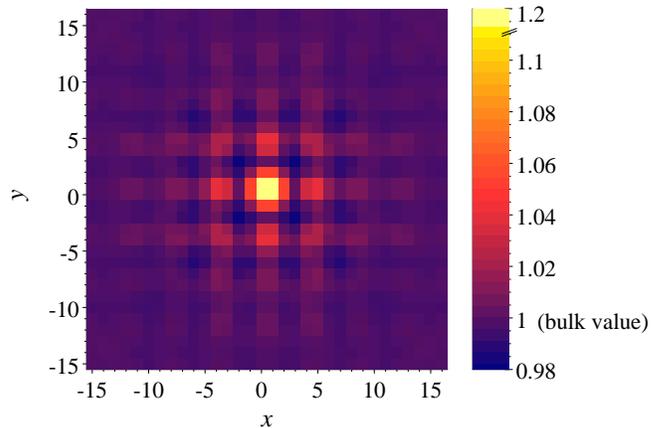}} \caption{As in
Fig.~\protect\ref{figldos1}, but now with a plaquette-centered
defect at ${\bf r}_0 = (0.5,0.5)$. } \label{figldos2}
\end{figure}

Examples of the spatial dependence of the LDOS of this model are
shown in Figs.~\ref{figldos1} and \ref{figldos2}.
Substantial Fourier components at ${\bf K}_{cx}$, ${\bf K}_{cy}$
were present at low energies both for positive and negative bias.

\begin{figure}
\epsfxsize=3in \centerline{\epsffile{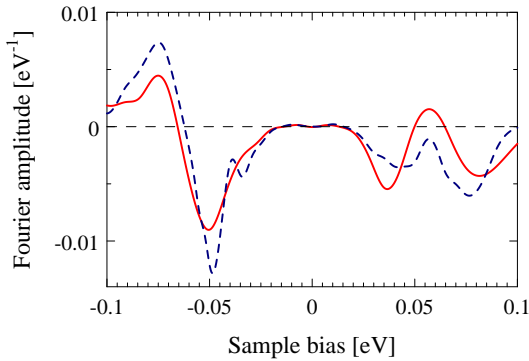}} \caption{As in
Fig~\protect\ref{tolyaft} but for the results obtained by summing
all the diagrams in Fig~\protect\ref{figdgr}. The solid line is
for the parameters in Fig.~\protect\ref{figldos1}, while the
dashed line is for $\gamma=0.14$~eV, $\zeta=0.2$~eV. This result
depends slightly on the system size (here 32$\times$32 sites) used
in the numerical calculation because of the slow decay of the LDOS
modulation. The peaks at $\approx \pm \Delta_0 = 0.04$~eV are
robust features of the theory, while the structures at other
energies depends upon the details of the band structure and
position of van Hove singularities.} \label{figen1}
\end{figure}

We also used the above results to compute (\ref{ft}) and the
results are shown in Fig.~\ref{figen1}.
A robust feature of the results in Figs~\ref{tolyaft} and
\ref{figen1} are the peaks in the modulus of the ${\bf K}_{cx,y}$
Fourier component near $\pm \Delta_0$. The remaining structure is
influenced by both the size of the spin gap, $\Delta$, and the
couplings $\zeta$, $\gamma$ (for a small range of $\zeta$ values,
multiple defect scattering processes can lead to a local
low-energy bound state in the SDW propagator), and depends
strongly on the fermionic host band structure, {\em i.e.\/}, the
particular form of $\varepsilon_{\bf k}$. Therefore we expect
different results for different compounds, but a detailed
comparison with experiment is difficult, as even the bare LDOS
seen in STM is not well represented by a two-dimensional
tight-binding form for $\varepsilon_{\bf k}$ -- in particular the
van-Hove singularity present in the theory is not seen
experimentally. There are also some differences between the two
models in Figs~\ref{tolyaft} and \ref{figen1} at small values of
the spin gap $\Delta$: these are due to the different pinning
terms near ${\bf r}_0$ discussed earlier, along with the
modification in the bulk fermion spectrum associated with the
first diagram on the r.h.s. of Fig~\ref{figdgr} which was included
only in Fig~\ref{figen1}.

This paper has presented a simple theory of the pinning of dynamic
SDW fluctuations. Using the dynamic spin susceptibility (as
measured in inelastic neutron scattering) and the quasiparticle
dispersion as input, we predict the energy dependence of the LDOS
at the charge ordering wavevector, as in Figs~\ref{tolyaft} and
\ref{figen1}. We hope that our method will ultimately allow a
quantitative comparison of the spectroscopy by inelastic neutron
scattering and by STM, and any possible disagreements should help
expose physical effects not covered by our simple model.

While this paper was being completed, two related preprints
appeared. Zhu {\em et al.} \cite{zhu} (like Ref.~\cite{ting})
discussed static magnetic order in the vortex core, in contrast to
our model of dynamic SDW. Howald {\em et al.} \cite{aharon}
reported STM measurements on Bi$_2$Sr$_2$CaCu$_2$O$_{8+x}$ in zero
field, including measurements of the energy dependence of the LDOS
at the wavevector $(\pi/2,0)$; the analogous quantity for the
charge order nucleated by an impurity is computed here in
Figs~\ref{tolyaft} and \ref{figen1}.

We thank Seamus Davis for a number of discussions on his STM
experiments \cite{seamus}, and for informing us about his method
of measuring (\ref{ft}). We are also grateful to E.~Demler,
A.~Kapitulnik, S.~Kivelson, and J.~Zaanen for useful discussions.
This research was supported by US NSF Grant DMR 0098226 (A.P. and
S.S.) and by the DFG through SFB 484 (M.V.).



\vspace{-0.1in}

\begin{thebibliography}{}

\vspace{-0.1in}

\bibitem{pwa} P.~W.~Anderson, Science, {\bf 235}, 1196 (1987).

\bibitem{vs} M.~Vojta and S.~Sachdev, \prl {\bf 83}, 3916
(1999); S.~Sachdev and N.~Read, Int. J. Mod. Phys. B {\bf 5}, 219
(1991).

\bibitem{vortex} S.~Sachdev, \prb {\bf 45}, 389 (1992);
N.~Nagaosa and P.~A.~Lee, \prb {\bf 45}, 966 (1992); D.~P.~Arovas
{\em et al.}, Phys. Rev. Lett. {\bf 79}, 2871 (1997); J.~H.~Han
and D.~H.~Lee, \prl {\bf 85}, 1100 (2000); M.~Franz and
Z.~Tesanovic, \prb {\bf 63},  064516  (2001); J.~I.~Kishine,
P.~A.~Lee, and X.-G.~Wen, \prl {\bf 86}, 5365 (2001).

\bibitem{ps} K.~Park and S.~Sachdev, \prb {\bf 64}, 184510
(2001).

\bibitem{seamus} J.~E.~Hoffman {\em et al.}, Science {\bf 295}, 466 (2002).

\bibitem{sdw} E.~Demler, S.~Sachdev, and Y.~Zhang, \prl {\bf
87}, 067202 (2001); Y.~Zhang, E.~Demler, and S.~Sachdev,
cond-mat/0112343.

\bibitem{white} S.~R.~White and D.~J.~Scalapino, \prl {\bf 80},
1272 (1998); \prb {\bf 61}, 6320 (2000).

\bibitem{jan} M.~Bosch, W.~van Saarloos, and J.~Zaanen, \prb {\bf 63},
092501 (2001); J.~Zaanen and A.~M.~Oles, Ann. Phys. (Leipzig) {\bf
5}, 224 (1996).

\bibitem{steve} S.~A.~Kivelson, E.~Fradkin, and V.~J.~Emery, Nature {\bf 393},
550 (1998); U.~Low, V.~J.~Emery, K.~Fabricius, and S.~A.~Kivelson,
\prl {\bf 72}, 1918 (1994).

\bibitem{troyer} H.~Tsunetsugu, M.~Troyer, and T.~M.~Rice, \prb
{\bf 51}, 16456 (1995).

\bibitem{castro} A.~H.~Castro Neto, Phys. Rev. B {\bf 64}, 104509
(2001).

\bibitem{assa} E.~Altman and A.~Auerbach,
cond-mat/0108087.

\bibitem{zachar} O.~Zachar, S.~A.~Kivelson, and V.~J.~Emery, \prb
{\bf 57}, 1422 (1998).

\bibitem{pphmf} A.~Polkovnikov, S.~Sachdev, M.~Vojta, and
E.~Demler, Proceedings of PPHMF IV, World Scientific, Singapore,
in press, cond-mat/0110329, computed LDOS modulations in a toy
model of electrons scattering off static charge density wave order
produced by the pinning of the sliding mode of the SDW. Here we
consider dynamic SDW fluctuations as the primary collective degree
of freedom and couple it directly to the electrons.

\bibitem{volovik} G.~E.~Volovik, Sov. Phys. JETP {\bf 58}, 469 (1993).

\bibitem{ting} Y.~Chen and C.~S.~Ting, cond-mat/0112369.

\bibitem{dispnote}
The numerical results were sensitive only to the SDW dispersion
$\omega_{\varphi_{x,y}}$ near the ordering wavevector, and not to
its behavior elsewhere.

\bibitem{zhu} J.-X.~Zhu, I.~Martin, and A.~R.~Bishop,
cond-mat/0201519.

\bibitem{aharon} C.~Howald {\em et al.},
cond-mat/0201546.

\end{thebibliography}
\end{document}